\documentstyle[twocolumn,seceq]{jpsj}

\def\runtitle{
One-Dimensional Electron Systems with 
$1/r^{\alpha}$ Interactions
}
\def\runauthor{
Yasumasa {\sc Tsukamoto} and Norio {\sc Kawakami}
} 
\title{
Critical Properties of 
One-Dimensional Electron Systems with $1/r^{\alpha}$-type 
Long-Range Interactions
}

\author{
Yasumasa {\sc Tsukamoto} and Norio {\sc Kawakami}
}

\inst{Department of Applied Physics,
Osaka University, Suita, Osaka 565-0871, Japan
}

\recdate{}

\abst{
We study the critical properties of one-dimensional electron 
systems with the $1/r^{\alpha}$-type long-range interaction 
($\alpha \ge 1$). Using bosonization methods, we discuss how the 
1D system exhibits a crossover from the Tomonaga-Luttinger  
liquid to the Wigner crystal when  $\alpha \rightarrow 1$. 
In particular,  a detailed analysis is given
to the Fermi-edge singularity phenomena when a {\it mobile} 
core hole is created suddenly in the valence band. It is found that 
the effect due to the core-hole motion on the Fermi-edge singularity 
is gradually suppressed as the system approaches the Wigner crystal, and
its critical behavior is dominated by  the ordinary impurity 
forward-scattering inherent in a static hole.
}
\kword
{long-range interaction, bosonization, Fermi-edge singularity}
\begin{document}
\sloppy
\maketitle
%%%%%%%%%%%%%%%%%%%%%%%%%%%%%%%%%
\section{Introduction}
%%%%%%%%%%%%%%%%%%%%%%%%%%%%%%%%%
Since the recent advances in technology made it possible to fabricate 
quantum wires, the study on one-dimensional (1D) 
electron systems has been one of the most attractive 
research areas in condensed matter physics. 
Low-energy properties of 1D correlated electron systems may be  
generally described by Tomonaga-Luttinger (TL) liquid theory.
\cite{haldane,emery,fukutaka,voire}
In this framework, the interaction among electrons 
is assumed to be short-ranged, which may result from  the 
cooperative screening effect. However, this assumption is not 
always valid for quasi-1D systems because the screening effect 
due to electrons becomes poorer in low-dimensional
systems, and thereby the long-range nature 
of the interaction may manifest itself in 
various quantities.  In fact, the optical experiments 
for quantum wires\cite{Goni} suggest that the effect of 
long-range interaction originating from the unscreened Coulomb force 
should be taken into account in treating 1D systems theoretically. 
More recently, it has been
recognized that  the unscreened long-range interaction 
is also important for quasi-1D  Carbon nanotubues.
\cite{Gogo,bal,yoshi}

Stimulated by these facts, many theoretical works on 1D 
electron systems with the $1/r$ Coulomb interaction have been reported. 
Among others, Schulz\cite{Schulz1} has clarified
that the $1/r$ long-range interaction enhances the 4$k_{F}$ 
charge density correlation,  driving 
the system to the {\it Wigner crystal} which shows 
a critical behavior quite different from the ordinary TL liquid.  
Subsequently, various properties in 1D systems 
with $1/r$ interaction have been examined.
\cite{Fabri,mau,Egg1,ogawa,yuan}
For instance,\cite{Fabri} 
a simple power-law temperature  dependence in the conductance 
$G$ for the TL liquid is changed to  $G \sim \exp [-\ln^{3/2}(1/T)]$, 
which vanishes with temperature faster than any power law, for the 
Wigner crystal with the $1/r$ interaction.  

In this paper, 
we investigate the critical behavior of 1D electron systems 
with the long-range interaction
 by means of abelian bosonization methods. 
To clearly see how the TL liquid is changed to
the Wigner crystal, we systematically deal with the $1/r^{\alpha}$-type 
long-range interaction with a continuously changing 
exponent $\alpha$. In particular, special attention 
is  focused on the Fermi-edge singularity
phenomena\cite{ogawa} with a {\it mobile} core hole.
We discuss the 
effect of the core-hole motion on the singularity
by computing the correlation functions for the 
orthogonality catastrophe (OC) and the Fermi-edge singularity (FES). 
It is found that the effect of a mobile core hole
is {\it suppressed} in the 1D system with long-range interaction, 
which is contrasted to our previous finding\cite{tsuka} that 
the effect of the core-hole motion plays an important 
role in TL liquids with short-range interaction. 

The paper is organized as follows. In \S2, we introduce 
the model with the  $1/r^{\alpha}$-type interaction, 
which also contains the impurity scattering due to 
a mobile core hole. Then in \S3, we  
 clarify how the  mobile core hole created
affects  the long-time critical properties in
the OC and the FES, by observing a crossover 
behavior from the TL liquid to the Wigner crystal.
A brief summary of the results is given 
in \S4.

%%%%%%%%%%%%%%%%%%%%%%%%%%%%%%%%%%%%%%%%%%%%%%%%%%%%%%
\section{Model with $1/r^\alpha$ Long-range Interaction}
%%%%%%%%%%%%%%%%%%%%%%%%%%%%%%%%%%%%%%%%%%%%%%%%%%%%%%
We consider a 1D correlated electron system with 
$1/r^\alpha$-type long-range interaction. 
As mentioned above,  the screening effect is not complete
in quasi-1D systems, so that the bare interaction is
modified from its pure Coulombic form in a 
non-trivial way. For instance, it has been suggested that for large 
distance the interaction might show the power-law behavior.\cite{power}
We shall consider here 
the $1/r^\alpha$ interaction with $\alpha \ge 1$, which 
can  systematically describe a crossover from  the TL liquid 
to the Wigner crystal.  
Since the $1/r^{\alpha}$ interaction exhibits a singular property 
for $r \rightarrow 0$, we introduce the interaction 
with short-range cut off $d$,\cite{Schulz1}
%%%%%%%%%%%%%%%%%%%%%%%%%%%%%%
\begin{eqnarray}
V(r) = \frac{1}{(r^2+d^2)^{\nu+1/2}}, {\hskip 10mm} \nu \geq 0,
\label{Vr}
\end{eqnarray}
%%%%%%%%%%%%%%%%%%%%%%%%%%%%%%
which is Fourier transformed as,
%%%%%%%%%%%%%%%%%%%%%%%%%%%%%%
\begin{eqnarray}
V(p) = \frac{2\sqrt \pi}{\Gamma(\nu+1/2)}
        \left| \frac{p}{2d} \right| ^{\nu} K_{\nu}(pd),
\label{Vp}
\end{eqnarray}
%%%%%%%%%%%%%%%%%%%%%%%%%%%%%%
where $\Gamma$ is the gamma function and 
$K_{\nu}$ denotes the $\nu$-th order modified Bessel function. 
Here we have introduced  the index $\nu$ ($\alpha=2\nu+1$), 
which is convenient for the following discussions.
By continuously changing $\nu$, we can alter the 
range of the interaction; For instance, in 
the  $\nu \rightarrow \infty$ limit 
we can effectively describe the $\delta$-function type 
interaction. For $\nu>0$ case, the uniform 
component of the interaction is given by,
%%%%%%%%%%%%%%%%%%%%%%%%%%%%%%%%%%%%%%%%%%%%%%
\begin{eqnarray}
V(0) =  \frac{\sqrt{\pi}\Gamma(\nu)}
{2d^{2\nu} \Gamma (\nu+1/2)} \equiv F(\nu,d)
\label{FG}
\end{eqnarray}
%%%%%%%%%%%%%%%%%%%%%%
which is the key quantity to control various 
correlation exponents so far as the system 
belongs to the TL liquid. 

The low-energy effective Hamiltonian for
the host part is now written down in the {\it g}-ology
description,\cite{haldane,emery,fukutaka,voire}
%%%%%%%%%%%%%%%%%%%%%%%%%%%%
\begin{eqnarray}
{\cal H}_{total} &=& {\cal H}_{\rho} + {\cal H}_{\sigma}\cr
{\cal H}_{\rho} &=& {\cal H}_{free}^{\rho} + {\cal H}_{g_2}^{\rho} 
                   + {\cal H}_{g_4}^{\rho},
\nonumber
\end{eqnarray}
%%%%%%%%%%%%%%%%%%
which consists of the charge ($\rho$) and spin ($\sigma$) sectors. 
The free part of the charge sector  has 
the ordinary harmonic form with respect to
the charge density $\rho_{+}(p) $ ($\rho_{-}(p)$)
for the right- (left-) going electrons, 
%%%%%%%%%%%%%%%
\begin{eqnarray}
{\cal H}_{free}^{\rho} &=&
\frac{\pi v_F}{L}
\sum_{p \neq 0}\sum_{r=\pm} :\rho_r(p)\rho_r(-p):  \cr
&+& \!\!
\frac{\pi}{2L}[v_{N_{\rho}}(N_{+\rho}+N_{-\rho})^2 \!\!
               + v_{J_{\rho}}(N_{+\rho}-N_{-\rho})^2 ]
\end{eqnarray}
%%%%%%%%%%%%%%%%%%%%%%%
where $v_F$ is the Fermi velocity and 
$v_{N_{\rho}}=v_{J_{\rho}}=v_F$. 
The long-range interaction defined by Eq.(\ref{Vp}) 
leads to  the forward scattering terms for the charge part,
%%%%%%%%%%%%%%%%
\begin{eqnarray}
{\cal H}_{g_2}^{\rho} &=& \frac{2}{L}
\sum_{p} g_2(p) \rho_+(p)\rho_-(-p)  \\
%%%
{\cal H}_{g_4}^{\rho} &=& \frac{1}{L}
\sum_{p,r=\pm} g_4(p):\rho_r(p)\rho_r(-p):
\end{eqnarray}
%%%%%%%%%%%%%%
where  $g_2(p)/\pi v_F=g_4(p)/\pi v_F=V(p)$.
Here we have neglected the backward and Umklapp scatterings, 
since they are not relevant in the present case.  
The spin part is given by SU(2)
critical  Hamiltonian,\cite{emery,fukutaka,voire} 
the explicit form  of which is not necessary in the
following treatment.

Let us now introduce the effective impurity scattering which describes 
the interaction between 1D electrons and 
a {\it mobile} core-hole 
created suddenly, 
e.g. by the X-ray absorption. In the previous paper,\cite{tsuka} 
we have derived the following 
effective Hamiltonian for the impurity scattering,
%%%%%%%%%%%%%%%%%%%%%%%%%%%%%
\begin{eqnarray}
{\cal H}_I^{\rho} &=& 
       \frac{v_f^{(1)}}{L}\sum_{p} \tilde{V}(p)[\rho_+(p) + \rho_-(p)] \cr
    &+&   \frac{v_f^{(2)}}{L}\sum_{p} [\rho_+(p) - \rho_-(p)].
\label{intH}
\end{eqnarray}
%%%%%%%%%%%%%%%%%%%%%%%%%%%%%%%%%%%%%%%%%%%%%
The first term is 
the ordinary forward scattering between electrons and an impurity,
for which we may change the spatial dependence of the impurity potential
by tuning the momentum dependent part $\tilde V(p)$.
The second nontrivial term does not exist
in the microscopic model, but is induced by the 
motion of the core hole; namely, it originates 
from the coupling between electron currents 
and the  core-hole motion.\cite{tsuka} 
Thus the interaction strength $v_f^{(2)}$
is dependent on the momentum of the mobile core hole.
For instance, if the impurity is assumed to be static(localized), 
then $v_f^{(2)}=0$.
This term, which is refereed to as the {\it asymmetric} forward 
scattering, plays an important role for FES in TL liquids. 
It was found\cite{tsuka} that the asymmetric 
scattering is to be short-ranged, even if the microscopic
interaction is long ranged.
In order to distinguish $v_f^{(1)}$ with $v_f^{(2)}$, 
we henceforth refer to the former as the {\it symmetric} 
forward scattering.

Following standard procedures in the 
bosonization,\cite{voire}
 ${\cal H}_{\rho}$ can be easily diagonalized.
Introducing the unitary operator 
$U$ ($U{\cal H}_{\rho}U^{\dagger} \equiv {\cal H}_{\rho}^{D}$), 
we have 
%%%
\begin{eqnarray}
 {\cal H}_{\rho}^{D}&=&\frac{\pi}{L}
\sum_{p \neq 0} \sum_{r=\pm}v_{\rho}(p):\rho_r(p) \rho_r(-p):\cr
 &+& \frac{\pi}{2L}
[v'_{N_{\rho}}(N_{+\rho}+N_{-\rho})^2
+ v'_{J_{\rho}}(N_{+\rho}-N_{-\rho})^2]
\end{eqnarray}
%%%%%%%%%%%%%%
where we define $v'_{N_{\rho}}=v_{\rho}/K_{\rho}$ and 
$v'_{J_{\rho}}=v_{\rho}K_{\rho}$
in terms of the TL parameter $K_{\rho}$ and the
charge velocity $v_{\rho}$ as, 
%%%%%%%%%%%%%%%%%%%%%%%%%%%%%%%
\begin{eqnarray}
K_{\rho}(p) = \frac{1}{\sqrt{1+2V(p)}},
v_{\rho}(p) = v_F \sqrt{1+2V(p)}.
\label{def Krho}
\end{eqnarray}
%%%%%%%%%%%%%%%%%%%%%%%%
The impurity potential ${\cal H}_I^{\rho}$ is 
simultaneously transformed as,
%%%%%%%%%%%%%%%%%%%%%%%%
\begin{eqnarray}
{\cal H}_{\rho}^{eff} 
      &\equiv& U {\cal H}_I^{\rho}U^{\dagger}  \cr
         &=& \frac{v_f^{(1)}}{L}\sum_{p} \tilde{V}(p)\sqrt{K_{\rho}(p)}
                          [\rho_+(p)+\rho_-(p)] \cr
 &+& \frac{v_f^{(2)}}{L}\sum_{p} \frac{1}{\sqrt{K_{\rho}(p)}}
                          [\rho_+(p)-\rho_-(p)].
\end{eqnarray}
%%%%%%%%%%%%%%%%%%%%%%%%%
Notice that the $p$ dependence of $K_{\rho}$ dramatically modifies 
the form of correlation functions.\cite{Schulz1,ogawa}

%%%%%%%%%%%%%%%%%%%%%%%%%%%%%%%%%%%%%%%%%%%%%%%%%%%%%%
\section{Correlation Functions with a Core Hole Suddenly Created}
%%%%%%%%%%%%%%%%%%%%%%%%%%%%%%%%%%%%%%%%%%%%%%%%%%%%%%

Let us now  discuss the critical properties for 
1D systems with $1/r^{\alpha}$ interaction. 
We here  focus on the FES phenomena with a {\it mobile} core hole
which is created in the valence band  by the X-ray 
absorption.\cite{tsuka,ogawa2} 
In particular, by analyzing the effects of 
the symmetric scattering $v_f^{(1)}$ and 
asymmetric scattering $v_f^{(2)}$ in Eq.(\ref{intH}), 
we clarify how these 
terms affect the correlation functions related to  the FES. 
We also  observe the crossover to the Wigner 
crystal  by continuously changing the power $\alpha$ in the interaction of
$1/r^\alpha$.  Note that the FES phenomena  
for the $1/r$ interaction have been discussed in detail
for a static core hole by Otani and Ogawa.\cite{ogawa}

To compute the correlation functions related to the FES, we 
introduce a core hole created suddenly 
at $t=0$, which thereby gives rise to the impurity potential
${\cal H}_I^{\rho}$. Then the resulting time-dependent correlation 
functions can be calculated by standard methods
\cite{mahan,shotte,ohtaka}. 
Here we employ the method suitable for our formulation based on  the
bosonization, i.e.  the 
{\it boundary condition changing operator}(BCCO) method.\cite{A&L}
The  BCCO is a kind  of the unitary transformations which is introduced 
so that the total Hamiltonian with impurity scattering
${\cal H}_{\rho}^{D}+{\cal H}_{\rho}^{eff}$ 
should be transformed into a free one. 
According to Affleck and Ludwig,\cite{A&L}
we can write down the BCCO $U_M(t)$ of our system as 
$U_M(t)=\exp[-iS_M(t)]$ with
%%%%%%%%%%%%%%%%%
\begin{eqnarray}
S_M(t) &=& \frac{\pi i}{L}\sum_{p\neq 0} \frac{e^{-\alpha |p|/2}}{p} \cr
       &\times& [ F_+(p)\rho_+(p)e^{ipv_{\rho}(p)t} 
            - F_-(p)\rho_-(p)e^{-ipv_{\rho}(p)t} ],
\end{eqnarray}
%%%%%%%%%%%%%%%%%%%%
where 
%%%%%%%%%%%%%%%%%%%%
\begin{eqnarray}
F_{\pm}(p) = 
     \frac{1}{\sqrt{2K_{\rho}(p)}} 
    \left(\frac{v_f^{(1)}}{\pi} \theta_1(p) \right)
      \pm 
     \sqrt{2K_{\rho}(p)} 
    \left(\frac{v_f^{(2)}}{\pi} \theta_2(p) \right).
\nonumber
\end{eqnarray}
%%%%%%%%%%%%%%%
Here $\theta_1(p)$ and $\theta_2(p)$ are given as,
%%%%%%%%%%%%%%%%
\begin{eqnarray}
\theta_1(p) &=& \frac{\sqrt{2}K_{\rho}(p)}{v_{\rho}(p)} \tilde{V}(p)
  = \frac{\sqrt{2}\tilde{V}(p)}{1+2V(p)} \equiv \Theta(p)\\
\theta_2(p) &=& \frac{1}{\sqrt{2}v_{\rho}(p)K_{\rho}(p)}
         =\frac{1}{\sqrt2} \equiv \theta,
\end{eqnarray}
%%%%%%%%%%%%%%%%%
where  we have used Eq.(\ref{def Krho}) to derive the second formulae.
Notice that since the symmetric scattering  
has $p$-dependence like $v_f^{(1)}\tilde{V}(p)$, the corresponding 
$\theta_1(p)$ remains $p$-dependent, whereas 
the asymmetric scattering $v_f^{(2)}$ is 
assumed to be of $\delta$-function type, 
so that $\theta_2(p)$ is independent of $p$.

By using the BCCO, we straightforwardly 
 write down the general formulae 
for the correlation functions of the orthogonality 
catastrophe (OC) and the Fermi-edge singularity (FES).
Making use of $\Theta(p)$ and $\theta$, 
the correlation function for 
the OC, $<U_M^{\dagger}(t)U_M(0)>$, is obtained as,
%%%%%%%%%%%%%%%%%%%
\begin{eqnarray}
&& Re<U_M^{\dagger}(t)U_M(0)> \cr
&=&  \!\!\exp \!\! \left[ \!
   - \!\! \left( \!\! \frac{v_f^{(1)}}{\pi} \!\! \right)^2
  \!\!\!      \int_{0}^{\infty}
\!\!\!\!\!
dp\frac{e^{-\alpha p}}{p}
            \frac{1}{4K_{\rho}(p)}\Theta(p)^2(1-\cos pv_{\rho}(p)t)
      \right]\cr
&\times& \!\!
  \exp \!\! \left[ \!
   - \!\! \left(\frac{v_f^{(2)}}{\pi} \theta \right)^2
   \!\!\!     \int_{0}^{\infty}
\!\!\!\!\!
dp\frac{e^{-\alpha p}}{p}
            K_{\rho}(p)(1-\cos pv_{\rho}(p)t)
      \right].\cr
\label{gOC}
\end{eqnarray}
%%%%%%%%%%%%%%%%%%%%%%%
This quantity characterizes how the density of
conduction electron responds when a mobile core hole
is suddenly created in the valence band. Note that the 
factor including 
$v_f^{(2)}$ is inherent in our system with a mobile core hole.
Similarly, the correlation function for the FES, for which
an electron is added to the conduction band can be derived as,
%%%%%%%%%%%%%%%%%%%
\begin{eqnarray}
&& Re<U_M^{\dagger}(t)\psi_r(0,t)\psi_r^{\dagger}(0,0)U_M(0)> \cr
 &=& \!\! \exp \!\! \left[ \!
   - \!\! \int_{0}^{\infty}
\!\!\!\!\!\!\!\!
dp\frac{e^{-\alpha p}}{p}
          \frac{1}{4K_{\rho}(p)} \!\!
      \left( \!\! 1 + \frac{v_f^{(1)}}{\pi} \Theta(p) \!\!\! \right)^2
        \!\!\!\!(1-\cos pv_{\rho}(p)t)
      \right]\cr
&\times& \!\!
     \exp \!\! \left[ 
   - \!\! \int_{0}^{\infty}
\!\!\!\!\!\!\!\!
dp\frac{e^{-\alpha p}}{p}
          K_{\rho}(p)
      \left( \frac{1}{2} + r \frac{v_f^{(2)}}{\pi}
        \theta \right)^2
        \!\!\!(1-\cos pv_{\rho}(p)t)
      \right] \cr
&\times& t^{-\frac{1}{2}},\cr
\label{gFES}
\end{eqnarray}
%%%%%%%%%%%%%%%
where the last term $t^{-\frac{1}{2}}$ is the contribution
from the spin sector with SU(2) symmetry.

Below, we discuss these correlation functions 
by explicitly evaluating the above integrals for each case,
for which  a mobile core hole is treated in several
circumstances. We first consider  the Wigner crystal 
with $1/r$ interaction as the host system, 
in  which two types of the impurity potential
are discussed: (a) $\tilde{V}(p)=V(p)$ 
and (b) $\tilde{V}(p)=1$. The former corresponds to
the impurity scattering by the $1/r$ potential, while the latter 
to the $\delta$-function potential.  To see the crossover to
the Wigner crystal, we then move to  the system with 
the $1/r^\alpha$ interaction ($\nu>0$) in Eq.(\ref{Vr}), and
calculate the correlation functions for two cases of impurity scattering, 
(a) $\tilde{V}(p)=V(p)$ case and (b) $\tilde{V}(p)=1$ cases. 
Note that the asymmetric forward 
scattering $v_f^{(2)}$  is always regarded to be short-ranged,
as mentioned in \S2.

%%---------------------------
\subsection{Wigner crystal: $V(r)=1/r$ case}
%%---------------------------

Let us begin with the case of 
the Wigner crystal in which the interaction between 
host electrons is of the $1/r$-type
($\nu=0$ in Eqs.(\ref{Vr}) and (\ref{Vp})). 

%%%%%%%%%%%%%%%%%%%%%%%%%%%%%%%%%%%%%%%%%%%%%%%
\subsubsection{$1/r$-type  impurity potential}
%%%%%%%%%%%%%%%%%%%%%%%%%%%%%%%%%%%%%%%%%%%%%%

Setting $\tilde{V}(p)=V(p)$, we first regard  the impurity scattering
due to a mobile core hole to be of the $ 1/r$ type.
Using the relation $V(p)=2K_0(pd)\simeq2|\ln|pd||\gg1$ for
small $p$, $\Theta(p)$ can be approximated as $\Theta(p)\simeq 1/\sqrt2$. 
Since the TL parameter $K_{\rho}$ still has the $p$-dependence,  
we evaluate Eq.(\ref{gOC}) by exploiting  the generalized formula 
derived by Otani and Ogawa,\cite{ogawa}
%%%%%%%%%%%%%%%%%%%
\begin{eqnarray}
&& Re<U_M^{\dagger}(t)U_M(0)> \cr 
&\simeq&
 \exp \left[ -\frac{1}{3} 
       \left( \frac{v_f^{(1)}}{\sqrt2 \pi} \right)^2 (\ln (t/d))^{3/2}
      \right]\cr
&\times&
 \exp \left[ -
       \left( \frac{v_f^{(2)}}{\sqrt2\pi} \right)^2 (\ln (t/d))^{1/2}
      \right].
\label{OC1} 
\end{eqnarray}
%%%%%%%%%%%%%%%%%

Similarly, we can derive the FES correlation function  as,
%%%%%%%%%%%%%%%%%%
\begin{eqnarray}
&& Re<U_M^{\dagger}(t)\psi_r(0,t)\psi_r^{\dagger}(0,0)U_M(0)> \cr 
&=& \exp \left[ -\frac{1}{3} 
     \left( 1+\frac{v_f^{(1)}}{\sqrt2 \pi} \right)^2 (\ln (t/d))^{3/2}
     \right]\cr
&\times&
 \exp \left[ 
      - \left( 
     \frac{1}{2} + r \frac{v_f^{(2)}}{\sqrt2 \pi} 
        \right)^2 (\ln (t/d))^{1/2}
      \right] t^{-1/2}.
\label{FES1}
\end{eqnarray}
%%%%%%%%%%%
It is seen that the first factor of $\exp[-{\rm const.}\ln^{3/2} (t/d)]$
in (\ref{OC1}) and (\ref{FES1}), which is caused by
the ordinary symmetric scattering
$v_f^{(1)}$, gives the most dominant contribution
to the correlation functions in the long-time regime. 
On the other hand, the asymmetric scattering inherent 
in the mobile core hole results in the subdominant factor,
 $\exp[-{\rm const.}\ln^{1/2}(t/d)]$. This 
implies that for the Wigner crystal  with $1/r$ interaction,
the effect of the core-hole motion is substantially suppressed 
due to the long-range nature of the host interaction.
 This conclusion is quite contrasted to the ordinary TL liquid case, 
for which both of the symmetric and asymmetric
scatterings equally contribute to the long-time behavior of 
FES phenomena.\cite{tsuka}  We note that  
our result Eq.(\ref{FES1}) 
for a static  core hole ($v_f^{(2)}=0$) 
agrees with those  obtained previously 
for the $1/r$ case.\cite{ogawa}

%%%%%%%%%%%%%%%%%%%%%%%%%%%%%%%%%%%%%%%%%%%%%%%%
\subsubsection{$\delta$-function impurity potential}
%%%%%%%%%%%%%%%%%%%%%%%%%%%%

The suppression of the impurity effect 
occurs more dramatically  when we consider the $\delta$-function
type impurity potential by setting $\tilde{V}(p)=1$. 
Such a problem with $\delta$-function scattering
in the Wigner crystal  has been 
treated previously for a static core 
hole by Fabrizio et al.\cite{Fabri}
Though $\Theta(p)$ has still $p$-dependence in this case, 
we can calculate the integrals in the 
same way mentioned above.
The OC correlation function takes the 
asymptotic form, 
%%%%%%%%%%%
\begin{eqnarray}
&& Re<U_M^{\dagger}(t)U_M(0)>\cr 
&=& 
  \exp \left[
            \frac{1}{4}\left(\frac{v_f^{(1)}}{\sqrt2 \pi}\right)^2
           (\ln(t/d))^{-1/2}
       \right]\cr
&\times&
  \exp \left[
             -\left(\frac{v_f^{(2)}}{\sqrt2 \pi}\right)^2
            (\ln(t/d))^{1/2}
       \right] \cdots
\end{eqnarray}
%%%%%%%%%%
%It is seen that the singular property is entirely 
%changed from the case with $1/r$ impurity 
%potential, although  the host interaction 
%is common for both cases.  We can see more 
%dramatic change in the FES correlation function, 
%which is given by,
It should be noted that in contrast to the result obtained 
in Eq.(\ref{OC1}), 
the asymmetric scattering $v_f^{(2)}$ inherent in a 
mobile core-hole plays the most dominant 
role for the OC in the long-time regime. 
However, we can see the dramatic suppression of the mobile impurity effect 
in the FES correlation function, which is given by,
%%%%%%%%%%%%%%%%%%%%%
\begin{eqnarray}
&& Re<U_M^{\dagger}(t)\psi_r(0,t)\psi_r^{\dagger}(0,0)U_M(0)> \cr 
&\simeq& 
   \exp \left[ - \frac{1}{3}  (\ln (t/d))^{3/2}
        \right]
\cdot 
   \exp \left[ -\left( \frac{v_f^{(1)}}{\sqrt2 \pi} \right)
               (\ln(t/d))^{1/2}
        \right]\cr
&\times&
   \exp \left[\frac{1}{4}\left( \frac{v_f^{(1)}}{\sqrt2 \pi} \right)^2
               (\ln(t/d))^{-1/2}
        \right]\cr
&\times&
   \exp \left[-\left(\frac{1}{2}+r\frac{v_f^{(2)}}{\sqrt2 \pi} \right)^2
               (\ln(t/d))^{1/2}
        \right] \times t^{-1/2} \cdots.
\label{FES2}
\end{eqnarray}
%%%%%%%%%%%%%%%%%%%%%
It is remarkable that the  impurity effect 
 disappears from the most dominant term
$\exp [- \frac{1}{3}  \ln ^{3/2}(t/d)]$; namely
neither of $v_f^{(1)}$ and $v_f^{(2)}$ 
contributes to the dominant term in the FES.
This peculiar  phenomenon 
is explained as follows. For the FES correlation function 
  an electron is added to the conduction band at $t=0$, 
which then gives rise to the $1/r$ potential 
scattering to the remaining conduction electrons.
This in turn causes the dominant singularity  
$\exp [- {\rm const.} \ln ^{3/2}(t/d)]$ in the 
long-time behavior, hiding the 
contribution from the core hole created.
This conclusion is consistent with 
the result deduced by Fabrizio et al.
for a static hole.\cite{Fabri} 

%------------------------------------
\subsection{Crossover to the 
Wigner crystal: $V(r)=1/r^{\alpha}$ case}
%------------------------------------

Now that the critical behavior in the Wigner crystal has been clarified,
let us study the TL liquid with  $V(r)=1/(r^2+d^2)^{\nu+1/2}$
interaction ($\nu>0$) in order to
see the crossover to the Wigner crystal.

%%%%%%%%%%%%%%%%%%%%%%%%%%%%%%%%%%%%%%%%%%%%%%%
\subsubsection{$1/r^{\alpha}$ impurity potential}
%%%%%%%%%%%%%%%%%%%%%%%%%%%%%%%%%%%%%%%%%%%%%%%%%

We first deal with the case of $\tilde{V}(p)=V(p)$;
i.e. the impurity potential takes the same 
form as that for the interaction among conduction 
electrons. From Eq.(\ref{FG}), one can see $V(p)\gg 1$ in the 
limit $\nu \rightarrow 0$, which 
results in  $\Theta(p)\simeq 1/\sqrt2$.
Therefore,  we end up with  $t^{-\alpha(\nu,d)}$-type {\it power-law} behavior 
with the exponents both for OC ($\alpha_{OC}(\nu,d)$) 
and for FES ($\alpha_{FES}(\nu,d)$), which  are obtained as,
%%%%%%%%%%%%%%%%%%%%%%%%%%%%%%%%%%%
\begin{eqnarray}
\alpha_{OC}(\nu,d) = \!\!
    \frac{1}{4} \!\! \left( \!\! \frac{v_f^{(1)}}{\sqrt2\pi} \!\! 
\right)^2
 \!\!\!   [2F(\nu,d)]^{\frac{1}{2}}
 + 
     \left( \!\! \frac{v_f^{(2)}}{\sqrt2\pi} \!\! \right)^2
 \!\!\!   [2F(\nu,d)]^{-\frac{1}{2}}\cr
\label{OC3}
\end{eqnarray}
%%%%%%%%%%%%%%%%%%
\begin{eqnarray}
\alpha_{FES}(\nu,d)\!\! &=& \!\!
  \frac{1}{4} \!\!
    \left( \!\! 1 + \frac{v_f^{(1)}}{\sqrt2 \pi} \!\! \right)^2 
 \!\!\!  [2F(\nu,d)]^{\frac{1}{2}}
 \! \cr &+&  \!\!\!
    \left( \!\! \frac{1}{2} + r\frac{v_f^{(2)}}{\sqrt2 \pi} \!\! \right)^2
  \!\!\!  [2F(\nu,d)]^{-\frac{1}{2}}
 + \frac{1}{2},
\label{FES3}
\end{eqnarray}
%%%%%%%%%%%%%%%%%%%%
where $F(\nu,d)$ is defined in  Eq.(\ref{FG}). 
Note that the above formulae for the critical exponents 
are typical for the TL liquid. Therefore, if the host interaction 
decays faster than $1/r$ (i.e. $\nu >0$), the FES correlation is  
governed by the power-law behavior, as should 
be the case for the TL liquid. In this case, both of the asymmetric 
scattering  $v_f^{(2)}$  and the symmetric scattering
$v_f^{(1)}$ contributes to the dominant power-law behavior.
Notice, however, that $F(\nu,d)$  becomes very large,
  in the limit of $\nu \rightarrow 0$,
which is seen from the expanded formula,
%%%%%%%%%%%%%%%%%%%%%%%%%%%%%%%%%%
\begin{eqnarray}
F(\nu,d) = \frac{1}{2d^\nu} \sum_{n=0}^{\infty} 
\frac{(2n-1)!!}{(2n)!!(\nu+n)}.
\label{FG1}
\end{eqnarray}
%%%%%%%%%%%%%%%%%%%%%%%%%%%%%%%%%%%%%
As a result, the first term with 
$v_f^{(1)}$ in Eqs. (\ref{OC3}) and  (\ref{FES3}) becomes more dominant 
than the  second terms with $v_f^{(2)}$. This implies that the symmetric
scattering becomes more  prominent as the system approaches 
the Wigner crystal.  In other words, the effect of
the core-hole motion, which emerges via the asymmetric
scattering $v_f^{(2)}$,  is dramatically suppressed 
for the FES in the limit  $\nu \rightarrow 0$,
though it plays the crucial role in the 
ordinary TL liquids.\cite{tsuka}  Finally at 
$\nu=0$, the most dominant term is replaced by 
$\exp[-{\rm const.} \ln ^{3/2}(t/d)]$  as shown in (\ref{FES1}), which is
typical for the long-time behavior due to the  static (localized) hole.

%%%%%%%%%%%%%%%%%%%%%%%%%
\subsubsection{ $\delta$-function impurity potential}
%%%%%%%%%%%%%%%%%%%%%%%%%

Even if the impurity potential is replaced by the one with
$\delta$-function type ($\tilde{V}(p)=1$), we 
find  that the correlation functions
show the power-law behavior inherent in TL liquids, whose 
exponents are given  by the Gaussian formula 
like  Eq.(\ref{OC3}) and Eq.(\ref{FES3}). 
Therefore,  as far as the system  stays in the  
TL liquid, the FES singularity is governed by the 
power-law behavior.  However, as mentioned in the previous
subsection, if we consider the Wigner crystal with 
$1/r$ interaction, there 
exist qualitative differences between the $1/r$ impurity
potential and $\delta$-function impurity potential.

%%%%%%%%%%%%%%%%%%%%%%%%%%%%%%%%%%%%
\subsection{Green function}
%%%%%%%%%%%%%%%%%%%%%%%%%%%%%%%%%%%%

To close this section, we briefly mention the long-distance behavior of
the one-particle Green function (GF), which was
 discussed by Schulz.\cite{Schulz1}
The long-distance critical behavior of the GF for $\nu>0$ (TL liquid) is  
simply obtained  by setting 
$v_f^{(1)}=v_f^{(2)}=0$ in the FES exponent (\ref{FES3}),
%%%%%%%%%%%%%%%%%%%%%%%%%%%%%%
\begin{eqnarray}
<\psi_r(x)\psi_r^{\dagger}(0)>
 = x^{-\alpha_G(\nu,d)}\cr
\alpha_G(\nu,d) = 
  \frac{1}{4} [ 2F(\nu,d) ]^{\frac{1}{2}}
 +  
  \frac{1}{4} [ 2F(\nu,d) ]^{-\frac{1}{2}}
 + \frac{1}{2}.
\label{GF for ALRI}
\end{eqnarray}
%%%%%%%%%%%%%%%%%%%%%%%%%%%%%
This exponent  takes a
typical form for the GF of the TL liquid,
%%%%%%%%%%%%%%%%%
\begin{eqnarray}
\frac{1}{4}\left( \frac{1}{K_{\rho}} +K_{\rho} \right) + \frac{1}{2}, 
\label{G-fn for TL}
\end{eqnarray}
%%%%%%%%%%%%%%%%%
with $K_\rho =[ 2F(\nu,d) ]^{-\frac{1}{2}}$.
On the other hand, by performing a similar procedure in 
Eq.(\ref{FES1}), 
we obtain the GF for the case of $1/r$ 
long-range interaction,
%%%%%%%%%%%%%%%%%%%%%%%%%%%%%%%%
\begin{eqnarray} 
&& <\psi_r(x)\psi_r^{\dagger}(0)> \cr 
&=& \exp \left[  -\frac{1}{3} (\ln (x/d))^{3/2}
         \right] 
 \exp \left[   -(\ln (x/d))^{1/2}
      \right] \cdots \cr &\times& x^{-1/2},
\label{GF for 1/r-type}
\end{eqnarray}
%%%%%%%%%%%%%%%%%%%%%%%%%%%%%%%%%%%%%%%%%%%%%%%%%%%%%%%%%
which reproduces the formula obtained by  
Schulz for the 1D Wigner Crystal.\cite{Schulz1}

It is now instructive to discuss the above GF 
in the viewpoint of a mobile core hole treated here, 
which allows us to clearly see the 
origin of the characteristic behavior in  the GF
for the Wigner crystal. Let us first notice that for $\nu>0$
the critical exponent (\ref{G-fn for TL})
for the charge sector has two contributions,  the one coming from
charged excitations, $1/K_{\rho}$, and the 
other from  the current excitations, $K_{\rho}$. By observing the
formula (\ref{FES3}), we can see that 
these should be referred to as the contributions from 
the symmetric and the  asymmetric 
impurity scatterings, respectively,
if we regard the added electron in the system  as a mobile impurity.
As $\nu \rightarrow 0$,
$1/K_\rho$ becomes very large, and finally at $\nu=0$ these two terms 
are respectively changed to the anomalous critical behaviors
$\exp[-\ln^{3/2}x]$ and  $\exp[-\ln^{1/2}x]$ 
inherent in the Wigner crystal.\cite{Schulz1} 
Therefore, we can naturally identify these anomalous 
contributions with those  from the charged (symmetric forward scattering)
and the current excitations (asymmetric forward scattering).
Based on these correspondences, we can say  that the reason 
why the term $\exp[-\ln^{1/2}x]$ is less dominant than
$\exp[-\ln^{3/2}x]$ is due to the fact that the 
asymmetric forward scattering  is to be short-ranged even if the 
the microscopic interaction is long-ranged, as mentioned 
in \S 2.

%%%%%%%%%%%%%%%%%
\section{Summary}
%%%%%%%%%%%%%%%%%

We have studied 1D electron systems with  
$1/r^\alpha$ long-range interaction to see how the TL
liquid exhibits a crossover to the Wigner crystal. 
In particular, we have focused on the FES phenomena 
with a {\it mobile} core hole suddenly created for the 
valence band, which contains the ordinary symmetric
forward scattering as well as the asymmetric
scattering. It has been shown that, if both of the interactions 
among the host electrons and the impurity potential
is of $1/r^\alpha$ type, 
the ordinary  symmetric scattering becomes more 
dominant than the  asymmetric scattering 
as the system approaches the Wigner crystal. 
This is quite contrasted to the ordinary TL liquid for which 
both of the symmetric and asymmetric potential scatterings play
an important role in the FES problem.
On the other hand, if the impurity scattering due to 
the mobile core hole are assumed to be short ranged while the
host electrons have the $1/r$ long-range interaction,  both of the 
symmetric and asymmetric scatterings, are entirely suppressed
 for the FES in  the Wigner crystal, 
being in contrast to the case with long-range impurity potential.

%%%%%%%%%%%%%%%%%
\acknowledgements
%%%%%%%%%%%%%%%%%

We would like to thank T. Fujii for valuable discussions.
This work was partly supported by a Grant-in-Aid from the Ministry
of Education, Science, Sports and Culture, Japan.


\begin{thebibliography}{99}
\bibitem{haldane}
 F. D. M. Haldane: J. Phys. C {\bf 14}, (1981) 2585;
Phys. Rev. Lett. {\bf 47}, (1981) 1840.
%
\bibitem{emery}
V.J. Emery: in {\it Highly Conducting One-Dimensional
Solids}, edited by J.T. Devreese {\it et al.}
(Plenum, New York, 1979).
%
\bibitem{fukutaka}
H. Fukuyama and H. Takayama: 
{\it Electronic Properties of
Inorganic Quasi-One-Dimensional Compounds,}
eds. P. Mon\c{c}eau (D. Reidel, 1985), p.41. 

\bibitem{voire}
J. Voit: Rep. Prog. Phys. {\bf 58}, (1995) 977.
%
\bibitem{Goni}
A.R. Go$\rm \tilde n$i, A. Pinczuk, J.S. Weiner, J.M. Calleja, B.S. Dennis, 
L.N. Pfeiffer, and K.W. West: 
Phys. Rev. Lett. {\bf 67}, (1991) 3298.
%
\bibitem{Gogo}
R. Egger and A.O. Gogolin: 
Phys. Rev. Lett. {\bf 79}, (1997) 5082.
%
\bibitem{bal}
C. Kane, L. Balents and M.P.A. Fisher: 
Phys. Rev. Lett. {\bf 79}, (1997) 5086.
%
\bibitem{yoshi}
H. Yoshioka and A.A. Odintsov: 
Phys. Rev. Lett. {\bf 82}, (1999) 374.
%
\bibitem{Schulz1}
H. J. Schulz: Phys. Rev. Lett. {\bf 71}, (1993) 1864.
%
\bibitem{Fabri}
M. Fabrizio, A. O. Gogolin, and Scheidl: 
Phys. Rev. Lett. {\bf 72}, (1994) 2235.
%
\bibitem{mau}
H. Maurey and T. Giamarchi: 
Phys. Rev. B {\bf 51}, (1995) 10833.
%
\bibitem{Egg1}
R. Egger and H. Grabert:
Phys. Rev. Lett. {\bf 75}, (1995) 3505.
%
\bibitem{ogawa}
H. Otani and T. Ogawa: 
Phys. Rev. B {\bf 54}, (1996) 4540.
%
\bibitem{yuan}
Q. Yuan, H. Chen Y. Zhang and Y. Chen: 
Phys. Rev. B {\bf 58}, (1998) 1084.
%
\bibitem{tsuka}
Y. Tsukamoto, T. Fujii and N. Kawakami: 
Phys. Rev. B {\bf 58}, (1998) 3633;{\it ibid.}
Eur. Phys. B {\bf 5}, (1998) 479.
%
\bibitem{power}
P. L.McEuen, E. B. Foxman, J. Kinaret, U. Meirav,
M. A. Kastner, N. S. Wingreen and S. J. Wind:
Phys. Rev. {\bf B45} (1992) 11419.

%%%%%%%%%%%%%%%%%%%%%%%%%%%%%%%%
\bibitem{ogawa2}
T. Ogawa, A. Furusaki and N. Nagaosa: 
Phys. Rev. Lett. {\bf 68}, (1992) 3638.
\bibitem{mahan}
G.D. Mahan: 
Phys. Rev. {\bf 153}, (1967) 882.
%
\bibitem{shotte}
K.D.Schotte and U. Schotte: 
Phys. Rev. {\bf 182}, (1969) 479.
%
\bibitem{ohtaka}
K. Ohtaka and Y. Tanabe: 
Rev. Mod. Phys. {\bf 62},(1990) 929 .
%
\bibitem{A&L}
I. Affleck and A.W.W. Ludwig: 
J. Phys. A. {\bf 27}, (1994) 5375.
%

\end{thebibliography}
\end{document}